\documentclass{jetpl}
\input epsf

\twocolumn
\lat
\title{Correlation function of ultra-high energy cosmic 
rays favors point sources.}
\rtitle{Correlation function of ultra-high energy\ldots}
\sodtitle{Correlation function of ultra-high energy cosmic 
rays favors point sources.}

\author{P.G.~Tinyakov$^{a,c}$ and I.I.~Tkachev$^{b,c}$ }
\rauthor{P.\,G.\,Tinyakov and I.\,I.\,Tkachev}
\sodauthor{Tinyakov, Tkachev}

\address{$^a${\it Institute of Theoretical Physics,\\
University of Lausanne, CH-1015 Lausanne, Switzerland\\}
$^b${\it CERN Theory Division, CH-1211 Geneva 23, Switzerland\\}
$^c${\it Institute for Nuclear Research of the 
Academy of Sciences of Russia, Moscow 117312, Russia }}

\abstract{We calculate the angular two-point correlation function of ultra-high
energy cosmic rays (UHECR) observed by AGASA and Yakutsk experiments.
In both data sets, there is a strong signal at highest energies, which
is concentrated in the first bin of the size of the angular resolution
of the experiment. For the uniform distribution of sources, the
probability of a chance clustering is $4\times 10^{-6}$. Correlations
are absent or not significant at larger angles. This favors the models
with compact sources of UHECR.}

\PACS{98.70.Sa}

\begin{document}

\maketitle

\paragraph*{1.}
The measurements of the flux of UHECR at energies of order
$10^{20}$~eV \cite{spectra} provide compelling
evidence of the absence of the Greisen-Zatsepin-Kuzmin (GZK) cutoff
\cite{GZK}. The resolution of this puzzle seems to be impossible
without invoking new physics or extreme astrophysics. All models
suggested so far can be classified in three groups, according to the
way the GZK cutoff is avoided: i) ``nearby source'' ii) weak
interaction with CMB iii) bump in the injection spectrum.

The possibility i) assumes that a substantial fraction of the observed
UHECR comes from a relatively nearby source(s) and thus is not subject
to the GZK cutoff. This idea may be realized in different ways,
examples being the models of decaying superheavy dark matter
\cite{shdm} or models in which UHECR emitted by nearby source(s) 
propagate diffusively in the galactic \cite{olinto} or extragalactic
\cite{piran} magnetic fields. Although models of this type generically
predict large-scale anisotropy \cite{Dubovsky:1998pu,piran}, they 
might still work.

In the option ii) the GZK cutoff is eliminated (or shifted to higher
energies) by assuming weak or non-standard interaction of primary
particles with the cosmic microwave background. This may happen, for
instance, if primary particles are neutrinos \cite{neutrinos},
hypothetical light SUSY hadrons \cite{kolb}, or if the Lorentz
invariance is violated at high energies \cite{VLI}. The
possibility iii) can be realized in models which involve topological
defects \cite{TD} or in some models where primary particles are
neutrinos \cite{Gelmini:2000ds}.

Existing data hint also at another important feature of
UHECR, namely, the clustering at small angles \cite{clusters1}. The
AGASA collaboration has reported three doublets and one triplet out of
47 events with energies $E>4\times 10^{19}$~eV, with chance
probability of less than 1\% in the case of the isotropic distribution
\cite{takeda}. The world data set has also been analyzed; 6 doublets
and 2 triplets out of 92 events with energies $E>4\times 10^{19}$~eV
were found \cite{uchihori}, with the chance probability less than 1\%.

If not a statistical fluctuation, what does the clustering imply for
models of UHECR?  There are two possible situations: either clustering
is due to the existence of point-like sources, or it is a result of
variations in the flux of UHECR over the sky (the regions
of higher flux are more likely to produce clusters of events
\cite{Medina-Tanco:2000rd}). In the first case, the models which
involve the diffuse propagation of UHECR are excluded. This case also
implies that there is no defocusing of UHECR in the magnetic fields
during their propagation. Such defocusing occurs even in a regular
(e.g., galactic) magnetic field, since different events in a cluster
have different energies. Thus, one can put bounds on the charge of the
primary particles. For the extra-galactic rays, knowing that primary
particles are charged would imply direct bounds on the extra-galactic
magnetic fields.

In the second case, the regions of higher flux may reflect
higher density of sources as in the models of superheavy dark matter
where they would correspond to dark matter clumps in the
halo. Alternatively, they may be due to the effects of propagation
such as defocusing in magnetic fields or magnetic lensing
\cite{lensing}.

In order to determine which of these two cases fits the present
experimental data better, it is not enough to know the probability to
have a certain number of clusters. In this respect previous analyses
\cite{clusters1,takeda,uchihori} are not sufficient. One
has to find the angular correlation function. This is the approach we
accept in this paper. 

\paragraph*{2.}
The two-point correlation function for a given set of events is
defined as follows. For each event, we divide the sphere into
concentric rings (bins) with fixed angular size (say, the angular
resolution of the experiment). We count the number of events falling
into each bin, sum over all events and divide by 2 to avoid double
counting, thus obtaining the numbers $N_i$. We repeat the same
procedure for a large number (typically $10^5$) of randomly generated
sets and calculate the mean Monte-Carlo value $N^{\rm MC}_i$ and the
variance $\sigma^{\rm MC}_i$ for each bin in a standard way. The 
correlation function can be defined as
$
f_i = {N_i/ N^{\rm MC}_i} - 1.
$
A deviation of $f_i$ from zero indicates the presence of the
correlations on the angular scale corresponding to $i$-th bin.

The correlation function $f_i$ fluctuates. In order to see whether its
deviation from zero is statistically significant, we define the ratio
$
r_i = ({N_i-  N^{\rm MC}_i)/ \sigma^{\rm MC}_i} , 
$
which shows the excess in the correlation function as compared to the
random distribution in the units of the variance. With enough
statistics, this quantity becomes a good measure of the probability of
the corresponding fluctuation.

The Monte-Carlo events are generated in the horizon reference frame
with the geometrical acceptance
$
dn \propto \cos \theta_z \sin \theta_z d\theta_z,
$
where $\theta_z$ is the zenith angle. Coordinates of the events are
then transformed into the equatorial frame assuming random
arrival time. We restrict our analysis to the events with zenith
angles $\theta_z < 45^{\circ}$ for which the experimental resolution
of arrival directions is the best \cite{uchihori}.

\begin{figure}
\centering\leavevmode\epsfxsize=3.4in\epsfbox{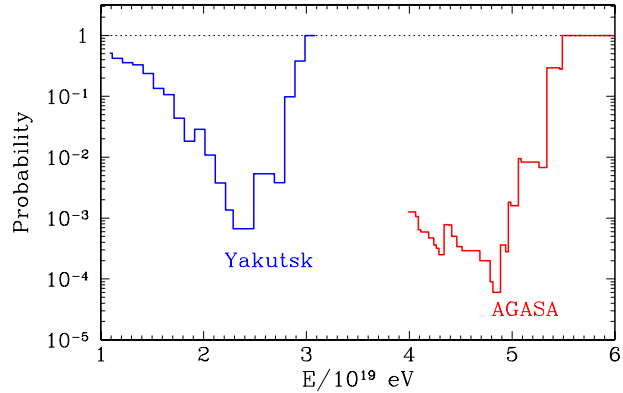}
\caption{FIG. 1. Probability to match or exceed the  observed count in the 
first bin as a function of energy for the random distribution of
arrival directions.}
\label{fig:Escans}
\end{figure}

If clusters at highest energies are not a statistical fluctuation, one
should expect that the spectrum consists of two components, the
clustered component taking over the uniform one at a certain
energy. The cut at an energy at which the clustered component starts
to dominate should give the most significant signal. Motivated by
these arguments, we calculated the probability of chance clustering as
a function of energy cut. We present here the results for the AGASA
\cite{takeda} and Yakutsk \cite{hp} data sets (other experiments are 
discussed in Section 3). For these simulations we took the bin size
equal to $2.5^{\circ}$ and $4^{\circ}$ for AGASA and Yakutsk,
respectively, which is the quoted (see e.g. \cite{takeda,uchihori})
angular resolution of each experiment multiplied by $\sqrt{2}$. The
results are summarized in Fig.~\ref{fig:Escans}, which shows the
probability to reproduce or exceed the observed count in the first
bin, as a function of the energy cut.
AGASA curve starts at $E=4\times10^{19}$~eV because the
data at smaller energies are not yet available. Yakutsk has much lower
statistics. Both curves rapidly rise to 1 in a similar way when the
statistics becomes poor. They suggest that the optimum energy cut is
higher than can be imposed at present statistics.

The difference between our results and those of Ref.~\cite{takeda}
(cf. Fig.~\ref{fig:Escans} of this paper and Fig.~12 of
Ref.~\cite{takeda}) is due to two reasons. First, 10 more events with
$E > 4\times 10^{19}$ have been observed \cite{Hayashida:1999zr} which
bring a new doublet. Second, and more important, we calculate a
different probability. The difference arises when there is a triplet
or higher multiplets in the data. In our approach a triplet is
equivalent to 3 or 2 doublets, depending on the relative position of
the events (compact or aligned), while higher multiplicity clusters
effectively have larger "weight". In Ref.~\cite{takeda} the
probabilities of doublets and triplets are calculated separately; the
probability of doublets is defined in such a way that a triplet
contributes as 3/2 of a doublet. The drawback of this method is that
the probabilities of doublets and triplets are not independent, and it
is not clear how to combine them. Triplets and higher multiplicity
clusters are better accounted for in our method, and the probability
of chance clustering which we get is lower than in Ref.~\cite{takeda}.
\begin{figure}
\begin{center}
\leavevmode\epsfxsize=3.4in\epsfbox{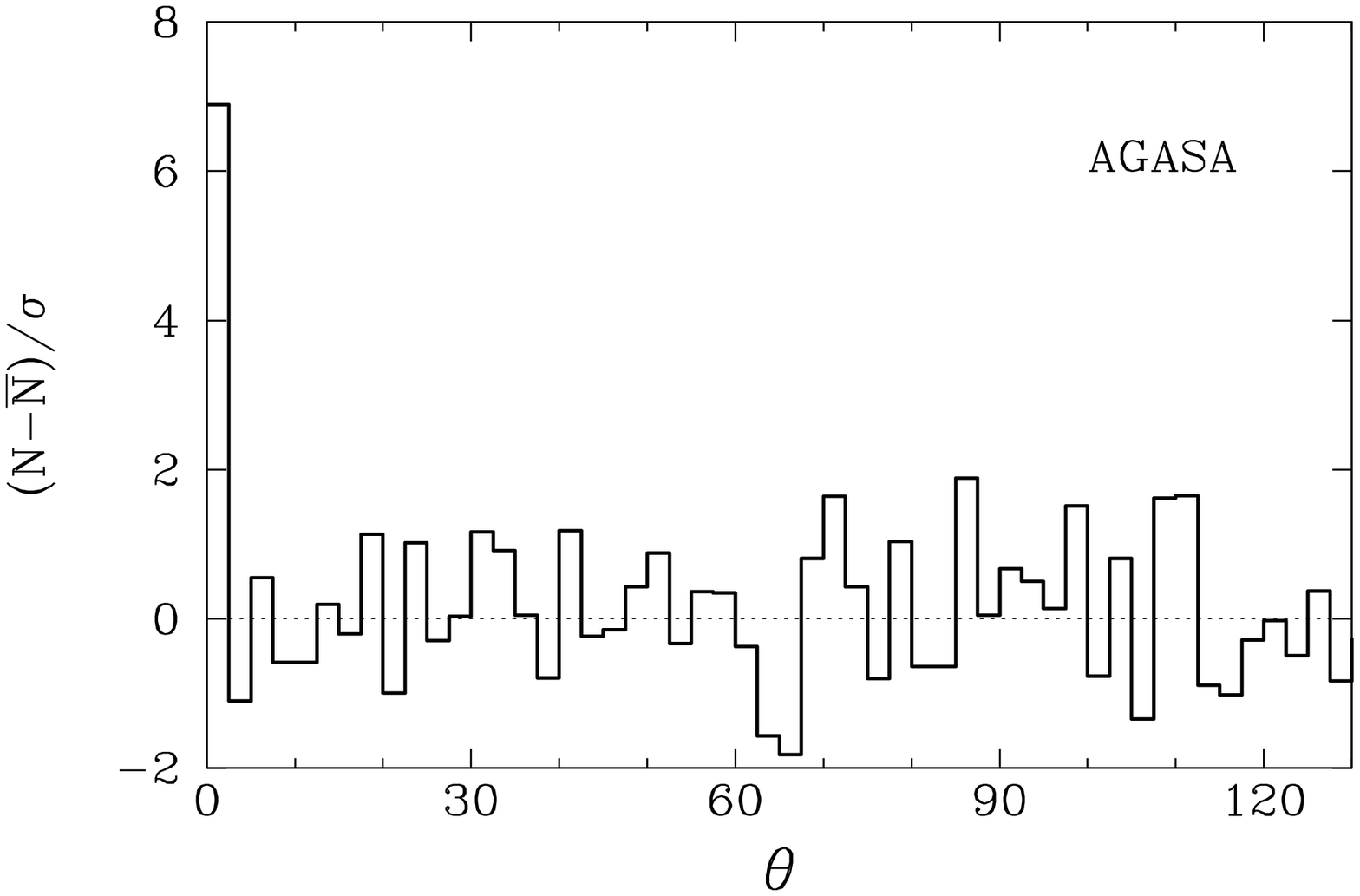}\\
\leavevmode\epsfxsize=3.4in\epsfbox{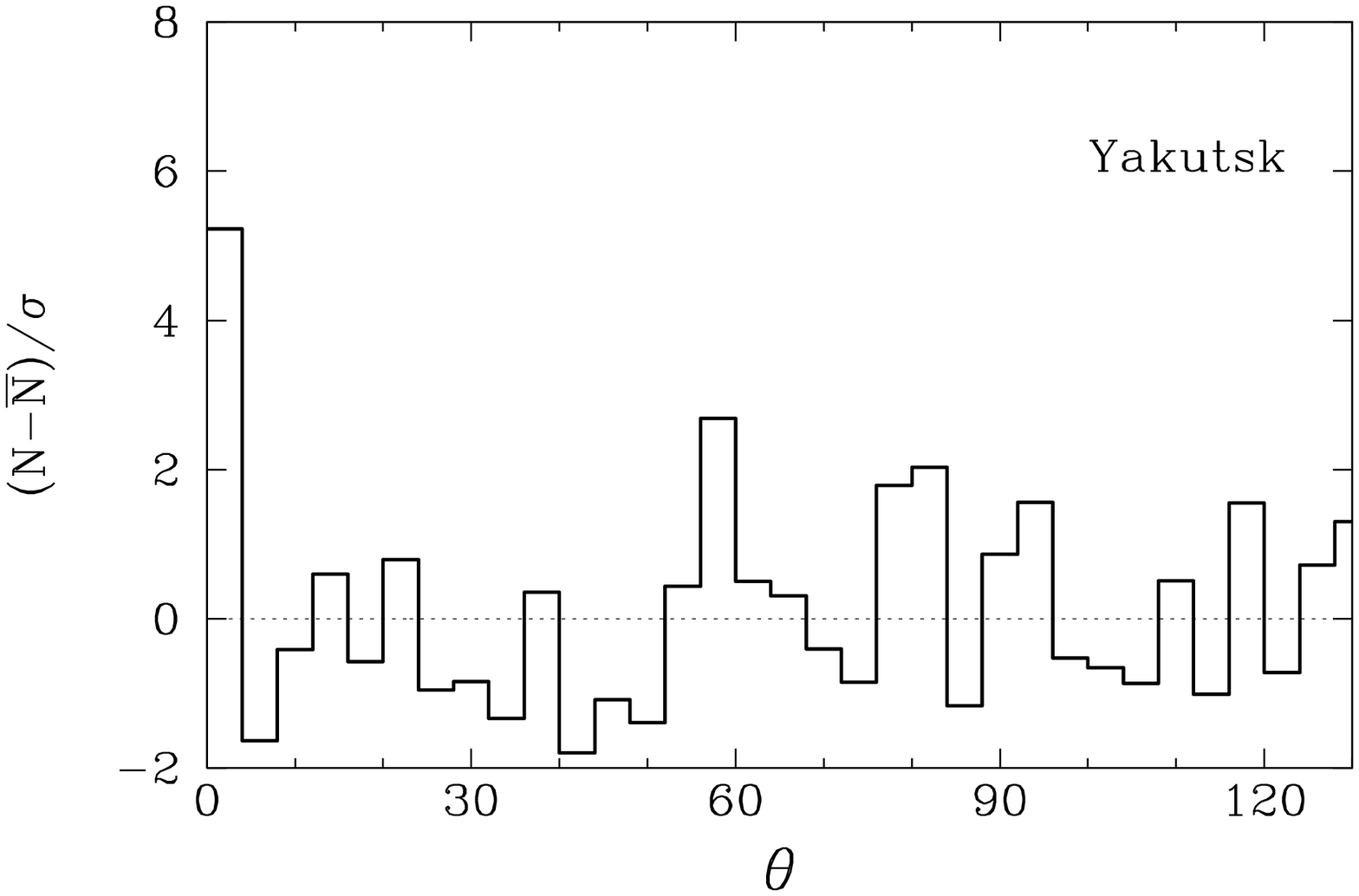}
\end{center}
\caption{FIG. 2. Angular correlation functions of UHECR with binning angles 
and cuts in energy quoted in the text.} \label{fig:CorrFun}
\end{figure}

Correlation functions calculated with the energy cuts corresponding to
the lowest chance probability is shown in Fig.~\ref{fig:CorrFun}.
Both AGASA and Yakutsk correlation functions have substantial excess
in the first bin.  The peak in AGASA curve corresponds to 6 doublets
(of which 3 actually form a triplet) out of 39 events. The peak in
Yakutsk curve corresponds to 8 doublets (of which 3 also come from a
triplet) out of 26 events.

Since the number of events in the first bin is not large, its
distribution is not well approximated by the Gaussian one, and the
deviation in the units of the variance is not a good measure for the
probability of fluctuations. We calculated the probability directly by
counting, in the Monte-Carlo simulation, the number of occurrences
with the same or larger number of events in the first
bin. Probabilities in the minima are small (see
Fig.~\ref{fig:Escans}), so we have recalculated them with $10^6$
Monte-Carlo sets.

As the lowest probabilities were obtained by scanning over the energy,
one may argue that they have to be multiplied by the number of steps
in the scan. This is not, however, correct because the results at
different energy cuts are not independent: higher energy set is a
subset of the lower energy one. As can be seen from
Fig.~\ref{fig:Escans}, the chance probability for AGASA is lower than
$10^{-3}$ in the whole energy range $(4-5)\times 10^{19}$~eV,
regardless of the number of steps in the scan. There still may be a
correction factor. To estimate it we made the following numerical
experiment. For $10^3$ randomly generated sets of events we have
performed exactly the same procedure as for the real data,
i.e. scanned over energies and obtained $10^3$ different minimum
probabilities. We found that the probability less than $10^{-2}$
occurred 27 times, while the probability less than $10^{-3}$ occurred
3 times. Thus we conclude that the correction factor is of order $3$.
This factor is included in the final results which are presented in
Table.1. 

\begin{table}
\caption{Table 1}
\begin{tabular}{c||c|c|c}
experiment &  bin size & $E_{\rm min}$ &  chance \\
&&&probability \\
\hline
AGASA  & $2.5^{\circ}$& $4.8\cdot 10^{19}$~eV & $3\cdot 10^{-4}$ \\
Yakutsk & $4^{\circ}$& $2.4\cdot 10^{19}$~eV &$2\cdot 10^{-3}$ 
\end{tabular}
\end{table}

\begin{figure}
\centering\leavevmode\epsfxsize=3.4in\epsfbox{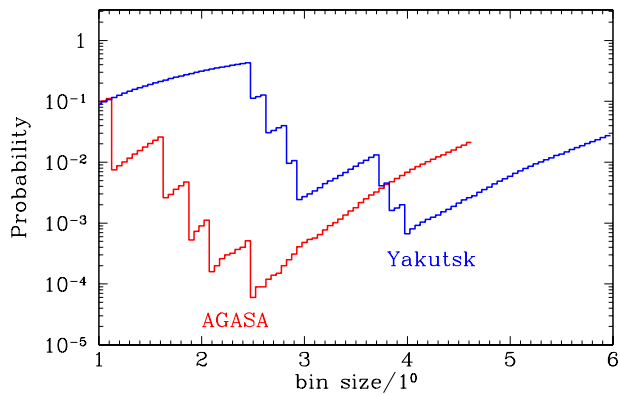}
\caption{FIG. 3. Probability to have observed count in the first bin as a 
function of the bin size. Cuts in energy correspond to minima
of Fig.~1.}
\label{fig:BinScan}
\end{figure}

We now turn to the determination of the angular size of the sources.
To this end we calculate the dependence of the probability to have the
observed (or larger) number of events in the first bin on the bin
size. This dependence is plotted in Fig.~\ref{fig:BinScan}. Jumps in
the curves occur when a new doublet enters the first bin. Despite
fluctuations, one can see that the minimum probability corresponds
roughly to $2.5^{\circ}$ and $4^{\circ}$ for AGASA and Yakutsk,
respectively. These numbers coincide with the angular resolutions of
the experiments, as is expected for sources with the angular size
smaller than the experimental resolution. Remarkably, there are no
doublets in the AGASA set with separations between $2.5^{\circ}$ and
$5^{\circ}$, while for the the extended source of the uniform
luminosity one would expect 4 times more events within $5^{\circ}$ as
there are within $2.5^{\circ}$. Thus, we conclude that the data favor
compact sources with angular size less than $2.5^{\circ}$.

If primary particles are charged, actual positions of sources differ
from the measured arrival directions because of the deflection in the
Galactic magnetic field (GMF). If the clustering is attributed to real
sources, it should not disappear but improve when the correction for
GMF is taken into account. We have simulated the effect of such
correction making use of the GMF models summarized in
Ref.~\cite{Stanev:1996qj}. For the charge $Z=1$ and BSS\_A model the
peak in the first bin does not change significantly; one cannot
discriminate between this case and the case of neutral particles. The
peak becomes small at $Z=2$, and disappears at larger $Z$ for all GMF
models of Ref.\cite{Stanev:1996qj}.

\paragraph*{3.} 
The other two UHECR experiments, Haverah Park (HP) and Volcano Ranch
(VR), do not see significant clustering \cite{HP}. With the energy cut
$E>2.4\times 10^{19}$~eV and the bin size $4^{\circ}$, the HP data
contain 2 doublets at 1.8 expected, while VR data contain 1 doublet at
0.1 expected with isotropic distribution. Let us estimate the combined
probability of clustering in all experiments assuming independent
Poisson distributions. The number of observed doublets in AGASA and
Yakutsk data are 6 and 8, respectively, while 0.87 and 2.2 are
expected (these ``effective'' expected numbers of doublets are
calculated from the condition that probabilities of Table~1 are
reproduced, i.e. ``penalty'' for the energy scan is included). Thus, 17
doublets are observed at 4.97 expected, which corresponds to the
Poisson probability $2\times 10^{-5}$. If HP data are excluded, the
probability becomes $1\times 10^{-6}$, while with both HP and VR data
excluded the probability is $4\times 10^{-6}$.

It is extremely unlikely that the clustering observed by AGASA and
Yakutsk experiments is a result of a random fluctuation in an
isotropic distribution. Rather, the working hypothesis should be the
existence of some number of compact sources which produce the observed
multiplets. Is this hypothesis consistent with HP and VR data?  For a
given experiment, the expected number of clusters is determined by the
total number of events \cite{dtt} (see also ref.\cite{fodor}); at
small clustering it scales like $N_{\rm tot}^{3/2}$ \cite{dtt}. Taking
AGASA data as a reference (6 doublets observed, 5.4 expected from
sources and 0.6 expected from chance clustering) allows to estimate
the expected number of doublets in other experiments by adding the
doublets expected from sources and the doublets expected from the
uniform background (calculated in the Monte-Carlo simulation). The
results are summarized in Table~2, together with corresponding Poisson
probabilities. 
\begin{table}
\caption{Table 2}
\begin{tabular}{c||c|c|c|c}
 & $N_{\rm tot}$  & observed & expected & probability \\
\hline
AGASA   & 39 & 6 & $5.4+0.6$    & $-$  \\
Yakutsk & 26 & 8 & $2.9+1.6$    & 0.09 \\
HP      & 32 & 2 & $4.0+1.8$    & 0.07 \\
VR      & 10 & 1 & $0.7+0.1$    & 0.55 
\label{t1}
\end{tabular}
\end{table}

All experiments are roughly consistent with the
assumption that the number of sources is such that they produce 5.4
doublets out of 39 events in average. Note that if HP data are
discarded \cite{HP}, the agreement between other experiments can be
made better.  

\begin{figure}
\centering\leavevmode\epsfxsize=3.6in\epsfbox{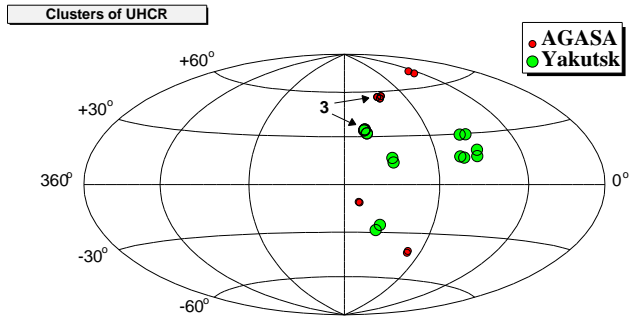}
\caption{FIG. 4. Observed clusters in Galactic coordinates. } 
\label{fig:Map}
\end{figure}

According to our simulations, the mean numbers of chance doublets are
0.6 and 1.6 for AGASA and Yakutsk, respectively. Therefore, most of
the clusters in AGASA and Yakutsk data are likely to be due to real
sources. In Fig.~\ref{fig:Map} we plot these clusters in the Galactic
coordinates (small and large circles correspond to AGASA and Yakutsk
events, respectively).  Positions of triplets are indicated by arrows.
The set of AGASA events with $E>4\times 10^{19}$~eV and Yakutsk events
with $E>2.4\times 10^{19}$~eV is a suitable choice for the search of
correlations with astrophysical objects. 

To summarize, the clustering of UHECR is statistically significant and
favors compact sources. This places further constraints on models
which can resolve the puzzle of the GZK cutoff. Those models which
involve large extragalactic magnetic fields, Ref.~\cite{piran}, as
well as models with heavy nuclei as primaries, e.g. \cite{olinto}, are
disfavored because they assume total isotropisation of original
arrival directions of UHECR. If violation of the Lorentz invariance is
the solution of the GZK puzzle, and primaries are protons, our results
place extremely strong limit on the extragalactic magnetic
fields. Regarding the models of decaying superheavy dark matter, it is
important to calculate \cite{CTT} the angular correlation function
predicted by these models and compare it to Fig.\ref{fig:CorrFun} in
order to see if the clumping on subgalactic scales can be responsible
for the clustering of UHECR.

\section*{Acknowledgments}
{\tolerance=400 We are grateful to S.I.~Bityukov, 
S.L.~Dubovsky, A.A.~Mikhailov, V.A.~Rubakov, M.E.~Shaposhnikov,
D.V.Semikoz and M.~Teshima for valuable comments and discussions. This
work is supported in part by the Swiss Science Foundation, grant
21-58947.99, and by INTAS grant 99-1065.  }


\begin{thebibliography}{99}
\bibitem{HP} 
We use the data of Ref.\cite{hp}. Unpublished events were added in the
analysis of Ref.\cite{uchihori} and more clusters were found.
\end{thebibliography}
\end{document}